\documentclass[10pt]{article}
\usepackage{psfig}
\usepackage{boxedminipage}
\usepackage{color}
\usepackage{multirow}
\renewcommand{\baselinestretch}{1.5}

\newcommand{\systemname}{{BODHI}}
\newtheorem{theorem}{Query}

\begin{document}

\title{The Building of BODHI, a Bio-diversity Database System}
\author{
\begin{tabular}{c c c c c c c c} 
{\bf Srikanta B J }\thanks{Contact Author: srikanta@dsl.serc.iisc.ernet.in} & &
{\bf Jayant R. Haritsa } & 
{\bf Uday S. Sen } &
\end{tabular}
\\ \\
\begin{tabular}{c}
Database Systems Lab, SERC\\
Indian Institute of Science \\
Bangalore~560012, India \\
\end{tabular}
}

\date{}
\maketitle

\begin{abstract} 

We have recently built a database system called BODHI, intended to store
plant bio-diversity information. It is based on an object-oriented modeling
approach and is developed completely around public-domain software. The unique
feature of BODHI is that it seamlessly integrates diverse types
of data, including \emph{taxonomic characteristics}, \emph{spatial
distributions}, and \emph{genetic sequences}, thereby spanning the
entire range from molecular to organism-level information. A variety
of sophisticated indexing strategies are incorporated to efficiently
access the various types of data, and a rule-based query processor is
employed for optimizing query execution. In this paper, we report on
our experiences in building BODHI and on its performance
characteristics for a representative set of queries.

\end{abstract}

\section{Introduction}
\label{sec:intro}

Over the last decade, there has been a revolutionary change in the way
biology has come to  be studied. Computer assisted experimentation and
data management have become commonplace in the biological sciences and
the branch of \emph{Bio-Informatics}  is drawing the attention of more
and more  researchers from  a variety of  disciplines.  A key  area of
interest here is the study  of the \emph{bio-diversity} of our planet.
The  database  research  community  has  also  realized  the  exciting
opportunities for  novel data management techniques in  this domain --
bio-diversity was featured  as the theme topic at  the Very Large DataBase
(VLDB) 2000 Conference \cite{DBLP:conf/vldb/LaneEN00}.

Over the  last three  years, we have  built a database  system, called
{\bf BODHI} (Bio-diversity Object Database arcHItecture)
\footnote{Gautama Buddha gained enlightenment under the Bodhi tree}, 
that  is  specifically designed  to  cater  to  the special  needs  of
biodiversity applications. While
BODHI currently hosts purely \emph{plant-related} data, it can be
easily extended  to supporting animal-related information  as well. In
this paper, we report on  our experiences in building BODHI, and
also present  its performance profile with regard  to a representative
set of user queries.

\subsection*{Background}
The  study   of  bio-diversity,  as   outlined  by  the   WCMC  (World
Conservation  Monitoring Center) \cite{wcmc-website}, is  an integrated
study of
\emph{Species}, \emph{Ecosystem} and \emph{Genetic} diversity. The data
associated  with these  domains vary  greatly  in the  scale of  their
structural  complexity, their  query processing  cost, and  also their
storage volume.  For example, while the  taxonomy information of
species diversity has complex hierarchical structure, spatial data
and  spatial  operators   associated  with  ecosystem  diversity  are
inherently  voluminous and  computationally expensive.   On  the other
hand, genetic  diversity is  based on specialized  pattern recognition
and similarity identification algorithms over DNA or Protein sequences
of the species.  Thus, supporting  such diverse domains under a single
integrated  platform  is a  challenge  to  the  data management  tools
currently  used  by  the  ecologists.   More  often  than  not,  these
scientists  make  use  of  \emph{different}  tools  for  managing  and
querying  over  each of  the domains,  leading  to difficulties  in
performing cross-domain queries.

To illustrate  the above point,  consider the following  target query,
which is of interest to  modern evolutionary biologists and similar to those that have appeared in the ecological literature(for example~\cite{ecoevol:92}):

\begin{theorem} 
Retrieve names of all fruit-bearing shrubs that share a part
of their  habitats and have a  Chromosomal DNA sequence  score of over
70 with Magnolia-champa.
\label{query:grand}
\end{theorem}

The above  query is typical in  the new age  of bio-diversity studies,
where  researchers  are simultaneously  studying  the macro-level  and
micro-level relationships  between various target  species.  Answering
the  query requires  the ability  to perform  integrated  queries over
taxonomy   hierarchies  (\emph{``fruit-bearing   shrubs''}),  recorded
spatial distribution  of species (\emph{``common  habitat''}), and the
genome  sequence  databases  (\emph{``Chromosomal DNA  sequence  score
above  70''}).  Unfortunately, however,  due to  the lack  of holistic
database  systems, biologists are  usually forced  to split  the query
into component queries, each of which can be processed separately over
independent databases, and then either \emph{manually} or through a
\emph{customized} tool perform the join of the results obtained from
the component queries.

For example,  a typical ``experience  story'' for answering  the above
query, as gathered from domain experts, would be:

\begin{enumerate}
\item Locate all fruit-bearing shrubs by performing a selection query
over the taxonomy database, stored in
\emph{MS-Access}~\cite{msaccess}, a ubiquitous PC-based relational database, and
retrieve the keys for their habitats.

\item For all the keys output in Step 1, access the associated habitat
data, stored  as polygons in  \emph{ArcView}~\cite{arcview}, a popular
spatial database product.  Then, for each qualifying polygon, find all
the  habitats in  the spatial  database that  intersect  this polygon.
Finally, compute an intersection  between the original set of polygons
and  the newly-derived  set of  polygons in  order to  prune  away the
habitats of organisms other than fruit-bearing shrubs.

\item From the output of Step 2, identify the names of the species of
the  target  shrubs,  and  then  perform  repeated  BLAST~\cite{blast}
searches  over  the  \emph{EMBL GenBank}~\cite{www:genbank}  DNA  sequence
database to  identify the sequences  (and, thereby the  species), that
have  a score  of  more than  70.   Note that  this final  score-based
pruning has to be performed externally by the researcher.

\end{enumerate}

\noindent
Long procedures, such as the above, for answering standard queries are
not  only cumbersome  but can  also  lead to  delays in  understanding
various micro-level and  macro-level bio-diversity patterns, and worse
yet,  the patterns  may  not be  found  at all  due  to limited  human
capabilities (example  to illustrate  this point can  be found  in the
field   of  molecular   biology,  reported   in~\cite{setubal},  where
comparison of sequences ``by hand'' missed out some of the significant
alignments  thereby   leading  to  erroneous   conclusions  about  the
functional similarity of proteins in question).

\subsection*{The BODHI System}
Based on  the above discussion, there  appears to be a  clear need for
building an \emph{integrated} database system that can be productively
used by  the bio-diversity community.   To address this need,  we have
built  the   \systemname~database  system  in   association  with  the
ecologists  and biologists  at our  institute.  The  project  has been
funded  by  the  Dept.~of   Biotechnology,  Ministry  of  Science  and
Technology, Government of India.

\systemname~is a \emph{native} object-oriented system that naturally
models the  complex objects ranging from hierarchies  to geometries to
sequences  that  are  intrinsic   to  the  bio-diversity  domain.   In
particular,   it  seamlessly  integrates   taxonomic  characteristics,
spatial  distributions, and  genomic sequences,  thereby  spanning the
range from  molecular to organism-level  information.  To the  best of
our knowledge,  \systemname~is the \emph{first system  to provide such
an integrated view}.

\systemname~is fully built around publicly available database components and
system software, and is  therefore completely free. In particular, the
SHORE  micro-kernel~\cite{shore94} from  the  University of  Wisconsin
(Madison) forms  the back-end of our software,  while the $\lambda$-DB
extensible  rule-based query  optimizer from  the University  of Texas
(Arlington) is  utilized for production of  efficient execution plans.
The system is currently  operational on a Pentium-III-based PC hosting
the Linux operating system.

A variety  of sophisticated access  structures, drawing on  the recent
research literature, have been implemented to provide efficient access
to     the     various    data     types.      For    example,     the
Path-Dictionary~\cite{pdindex98}        and       Multi-key       Type
indexes~\cite{mtindex97}   accelerate   access   to  inheritance   and
aggregation  hierarchies,  while  the  R*-tree~\cite{rstartree90}  and
Hilbert  R-tree~\cite{hilberttree} are  used  for negotiating  spatial
queries.

The BODHI server is  compliant with the ODMG standard~\cite{cattel93},
supporting an  OQL/ODL query and  data modeling interface.   To enable
biologists to interface with the system in a more intuitive manner,
\systemname~also supports access through the Web client-server model wherein
clients submit requests through the HTTP protocol and CGI-bin scripts,
and the results are  provided through the browser interface.  Further,
the server  is ``XML-friendly'', outputting the result  objects in XML
format, enabling  clients to visualize  the results in  their favorite
metaphor.

We view BODHI's role as not merely that of a database system in isolation,
but as a central repository that provides a common information exchange
platform for all the tools used in a biologist's ``data workbench'' such
as decision support systems, visualization packages, etc.  That is,
BODHI occupies a role similar to that played by the Management Information
Base (MIB) in tele-communication network management.

Algorithms proposed in the research literature  typically tend to
be evaluated in isolation and it is never clear whether their claimed
benefits really carry through in practice with regard to end-user metrics
in complete systems.  We suggest that researchers may find it possible
to address this deficiency by using BODHI as a ``test-bed'' on which
new ideas can be evaluated in a real-world kind of setting. As reported
later in this paper, we have ourselves carried out this exercise with
regard to spatial indexes.

Finally, BODHI is living proof that developing a viable biological DBMS
does not necessarily entail expensive hardware or software but can be
cobbled together using commodity components.

In this paper,  we report on our experiences  in building \systemname,
and   also  present  its   performance  profile   with  regard   to  a
representative set of biological  queries (including Query 1 mentioned
above).%
\footnote{A preliminary position paper focusing solely on the BODHI
architectural  design was presented  in \cite{comad2000}.}   Since, as
mentioned earlier, there  are no comparable systems that  we are aware
of, for  the most part our results  can be placed only  in an absolute
perspective.  However,  specifically for queries  restricted solely to
spatial  data, we  were able  to utilize  the well-known  Sequoia 2000
benchmark~\cite{sequoia2000},   and   additional   spatial   aggregate
operators such as  \emph{Closest} introduced in the~\cite{paradise94}.
Here our numbers  are competitive with those obtained  by the Paradise
GIS system~\cite{paradise94},  that was highly  optimized for handling
only spatial queries.

\subsection{Organization}

The remainder of  the paper is organized as  follows: Desirable design
goals for bio-diversity DBMS  are laid out in Section~\ref{sec:goals}.
The BODHI  system architecture and  its implementation are  covered in
Section~\ref{sec:design}  and~\ref{sec:impl},  respectively.   Then in
Section~\ref{sec:experience}, we  present our experiences  in building
BODHI,  and  followup  with   a  detailed  performance  evaluation  in
Section~\ref{sec:expt}.      Related    work     is     reviewed    in
Section~\ref{sec:related}.   Finally,  in Section~\ref{sec:concl},  we
present our conclusions and future research avenues.

\section{Design Goals}
\label{sec:goals}
In this section, we highlight the main features that would be
desirable in a bio-diversity information system. These include:
\subsection{Handling of Complex Data Types}
Plant bio-diversity data can be broadly classified into the following
three categories: 
\begin{description}
\item[Taxonomy Data]
This is data about the relationships between species based on their
characteristics. This includes \emph{phenetic relationships} (based
on comparison of physical characteristics) and \emph{phylogenetic
relationships} (based on evolutionary theory)\cite{pankhurst91}. The
various characteristics on which these relationships depend may vary
in time due to discovery of a new class of characteristics, corrections
to previously recorded characteristics, etc.

\item[Geo-spatial Data]
The study of ecology of species involves recording the geographical
and geological features of their habitats, water-bodies, artificial
structures such as highways which might affect the ecology, etc.  These are
represented on a map of the region and have to be handled as spatial
data by the database.

\item[Bio-molecular Data]
The genetic makeup of species is becoming increasingly important with
a large number of genome sequencing projects working on organisms and
plants.  For example, ``bio-prospectors'' look for indigenous sources
of medicines, pesticides and other useful extracts. Such data can be
discovered from the biomolecular and genetic composition of species.
\end{description}

The  above  datatypes  have  complex and  deeply-nested  relationships
within and between themselves. Further, they may involve sophisticated
structures such as sequences and sets.

\subsection{Molecular Pattern Discovery}
The  molecules that  are  of  interest in  bio-diversity  are DNA  and
Proteins.   DNA is  represented as  a long  sequence based  on  a four
nucleotide alphabet. There are regions in the DNA sequence, called
\emph{exons}, which contain the genetic code for the synthesis of
Proteins. The proteins are long chains of 20 amino acids. Each protein
is characterized by the amino acid patterns it has, and is responsible
for   various  functionalities   in   a  cell   which  determine   the
characteristics of the organism or plant.

The similarity  between two  genetic sequences is  a measure  of their
functional  similarity. Analysis  of  DNA and  Protein sequences  from
different  sources  gives  important  clues about  the  structure  and
function  of proteins,  evolutionary relationships  between organisms,
and helps in discovering drug targets.

There are a number of popular algorithms, such as Dynamic Programming,
BLAST~\cite{blast},   FastA~\cite{fasta}  etc.,   for   performing  the
similarity   search   over   genetic   sequences.    Researchers   and
bio-prospectors frequently search  the database using these algorithms
to locate gene sequences  of interest.  However, the implementation of
these algorithms  is typically external  to the database,  making them
relatively  slow.  It  therefore  appears attractive  to consider  the
possibility of integrating these algorithms in the database engine (this observation is gaining currency in the commercial database arena as well, as  exemplified by IBM's provision of homology searching through UDFs in DB2).

\subsection{Usage Interface}
As with all other  scientific communities, the bio-diversity community
relies  on  timely  knowledge  dissemination.   Therefore,  supporting
access through the Internet is vital for maximizing the utility of the
information stored in the database.

Typically, bio-diversity data is autonomously collected and managed by
individual research institutions  and commercial enterprises. In order
to provide  larger scope  of data availability,  it is  necessary that
such localized  and autonomous data  repositories be able  to exchange
data.   The  current state  of  information  exchange amongst  various
bio-diversity       data      repositories      is       not      very
satisfactory\cite{oecd:gbif}.  However,  with the advent  of XML, many
research groups are proposing DTDs in individual fields of ecology and
genetics\cite{anzmeta, bioml} .   The bio-diversity information system
should support these  DTDs for handling data over  heterogenous set of
repositories.

It is imperative to have a  good visualization interface for the
results produced by the system since (a) the end-users are biologists, not computer scientists
and (b) the results could range from simple text to multidimensional
spatial objects.

Finally, most of the research  in bio-diversity is done by small teams
of researchers  who work within low  budgets and are  unable to afford
high-cost  data  repository systems.   Therefore,  solutions that  are
completely  or largely based  on public-domain  freeware which  can be
hosted on  commodity hardware, with  total cost not  exceeding \$1000,
are essential for these groups.

\section{Architectural Overview Of BODHI}
\label{sec:design}
As  mentioned earlier, bio-diversity  data is  inherently hierarchical
and  has  complex relationships.  In  order  to enable  \emph{natural}
modeling  of these  entities and  their relationships,  \systemname~is
designed as an
\emph{object oriented} database server, with OQL/ODL query and data
modeling interfaces. While we consciously adopted this technology from
the very  beginning of our project  in 1998, it is  gratifying to note
that the  same approach is  now being taken by  large-scale biological
repositories such  as EMBL (European Molecular  Biology Laboratory) --
in a recent report, they  have indicated their intention in moving from
their   current  Oracle-based   relational  database   system   to  an
object-based data management and distribution scheme for their massive
genomic databases~\cite{embl:corba2000}.

\begin{figure*}
\begin{center}
\centerline{\psfig{file=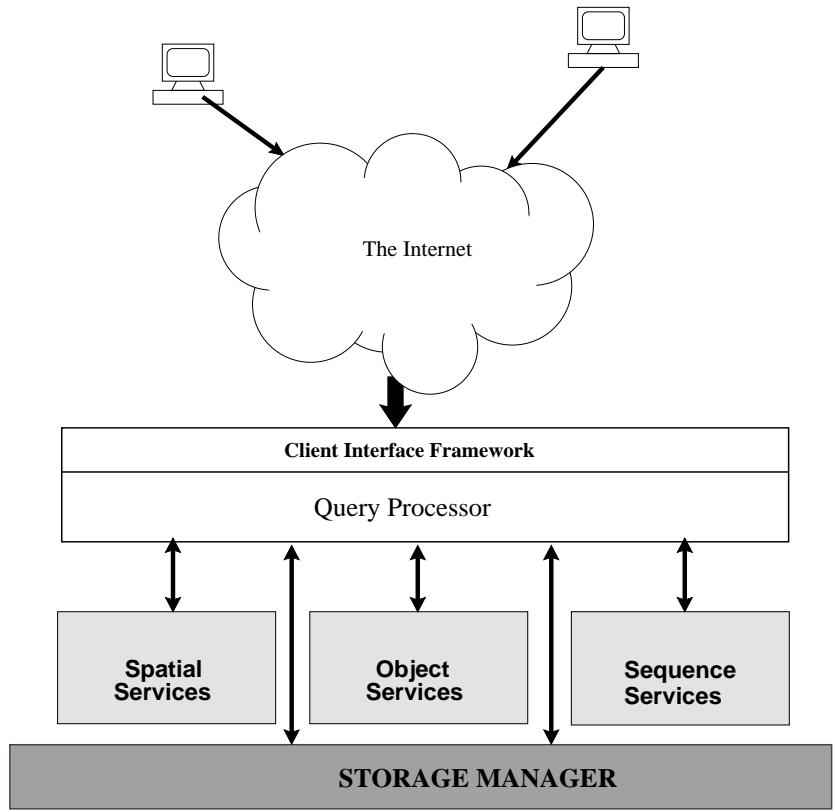,width=11cm}}
\caption{Schematic of Architecture of \systemname}
\label{fig:arch}
\end{center}
\end{figure*}

The    overall    architecture     of    \systemname~is    shown    in
Figure~\ref{fig:arch}.   At the  base  is the  storage manager,  which
provides the fundamental needs of a database server such as device and
storage  management,  transaction  processing,  logging  and  recovery
management.   The  application-specific   modules,  which  supply  the
taxonomic, spatial  and genomic services, are built  over this storage
manager  and  form the  functional  core  of  the system.   The  query
processor interfaces  with the  functional modules and  performs query
processing  and  optimization   using  statistics  exported  by  these
modules. Finally, the client  interface framework receives query forms
over  the Internet  from clients  and returns  results in  the desired
format.   In the  remainder  of  this section,  we  describe the  core
database components in more detail.

\subsection{Service Modules}

The  three  service  modules:  \emph{Object  Services},  \emph{Spatial
Services}  and \emph{Sequence  Services}, provide  the functionalities
for  each   of  the  bio-diversity  data  domains   mentioned  in  the
Introduction:

\subsubsection*{Object Services}

While  the storage  manager  handles basic  object  management, it  is
necessary  to   support  specialized  access   methods  for  efficient
processing of  queries over the  object schema and  its instantiation.
The  \emph{Object Services}  component bundles  together  these access
methods.

In  querying  over object  oriented  data  models,  it is  common  for
predicates   to   follow   arbitrarily  long   (sometimes   recursive)
relationship  paths, or  be  evaluated over  an inheritance  hierarchy
rooted at a chosen base type. As illustrations, consider the following
query types over a typical bio-diversity data model such as that given
in Figure~\ref{fig:model}:

\begin{figure*}[htb]
\begin{center}
\centerline{\psfig{file=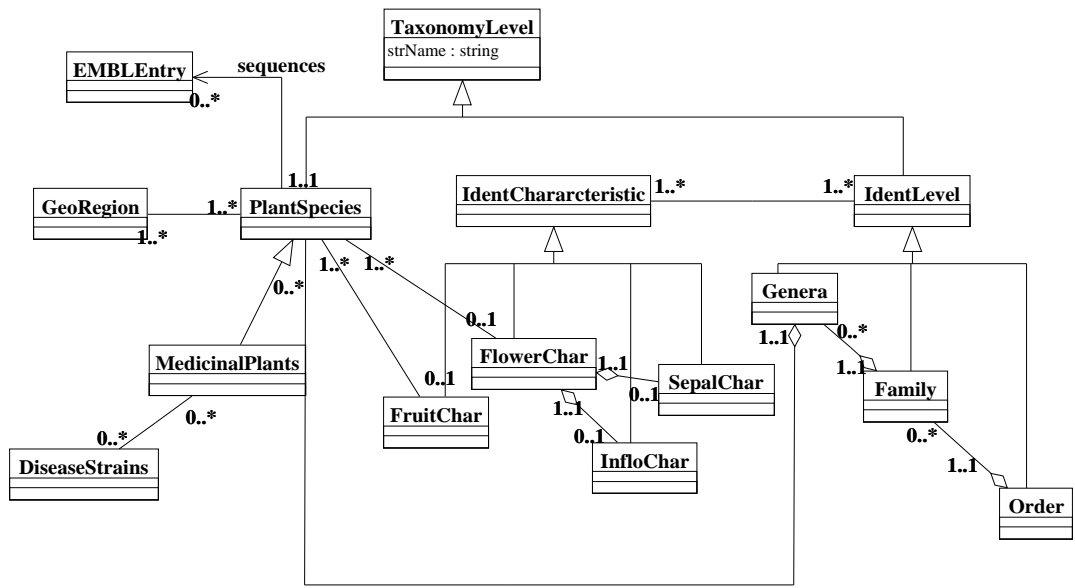}}
\caption{Partial Objectmodel}
\label{fig:model}
\end{center}
\end{figure*}

\begin{enumerate}
\item 
\emph{Identify the {\bf PlantSpecies} based on one or more of 
its {\bf IdentCharacteristics}.}
\item
\emph{Retrieve all {\bf IdentCharacteristics} of a given {\bf PlantSpecies}.} 
\item
\emph{List the names of all {\bf PlantSpecies} associated with a {\bf GeoRegion}.} 
\end{enumerate}

The above  queries illustrate the fact that  queries over relationship
graphs of bio-diversity data models  may have either an ancestor class
or a  nested class as  the predicate, and  might need to  be evaluated
over a inheritance hierarchy.  These queries may involve joins between
extents of objects  in the traversal paths, or  scanning over multiple
extents  for  the   predicate  in  the  case  of   queries  over  type
hierarchies.   Therefore,  access  methods  for both  inheritance  and
aggregation hierarchies are included in this module.

\subsubsection*{Spatial Services}
\begin{figure*}
\begin{center}
\centerline{\psfig{file=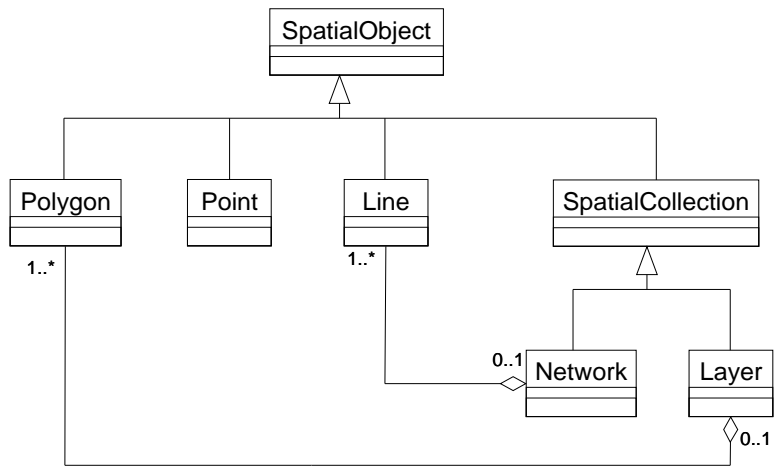,width=4in}}
\caption{Class diagram of Spatial Data model in \systemname}
\label{uml:spat}
\end{center}
\end{figure*}

Spatial  (or geographic)  data,  in both  vector  (object) and  raster
(bitmap)   formats,  constitutes   the  bulk   of   the  bio-diversity
information.   Due to  the inherent  complexity of  spatial operations
(such  as \emph{overlap}, \emph{closest},  etc.), combined  with large
volumes of data, spatial query  processing is considered to be a major
bottleneck in the expeditious processing of a cross-domain query (such
as Query~\ref{query:grand} in the Introduction).

The \emph{Spatial Services}  module provides efficient implementations
of access methods and spatial operations.
To ensure that the access methods have efficient disk allocations,
and thereby alleviate the performance bottleneck mentioned above,
these methods are built  \emph{within} the storage manager. While this
choice makes it cumbersome to replace or  upgrade the storage manager,
we felt that the resulting performance benefits would outweigh the
disadvantages.  

The Spatial module provides a spatial type system based on the
\emph{ROSE Algebra}~\cite{guting94}.  These types consist of
\emph{Simple primitives}: Point, Polyline, and Polygon; and
\emph{Compound primitives}: Layer and Network, which are collections of
related  Polygons  and  Polylines,  respectively.   The  spatial  type
hierarchy     supported    is     this    module     illustrated    in
Figure~\ref{uml:spat}.

\subsubsection*{Sequence Services}
In  modern  bio-diversity studies,  genetic  data  plays an  important
role~\cite{ecoevol:92}.   The \emph{Sequence Services}
module  interfaces  with  the  storage
manager  to  provide  efficient   storage  of  genetic  sequences  and
sequence retrieval  algorithms such  as BLAST,  FasTa, etc.  These
algorithms  are expensive  to  compute since  there  are currently  no
obvious ways of caching or indexing to speed up their computation, and
a full scan of the sequence database is therefore entailed each time.  The
Sequence Services module uses appropriate storage structures
for efficient execution of the genetic algorithms.

This module supports two primitive types: \emph{DNA} and
\emph{Protein}. The DNA alphabet of 4 nucleotides is encoded using two
bits and similarly  the Protein sequence alphabet of  20 is encoded in
five bits. The functions for translation of DNA sequences into Protein
sequences,  and vice-versa, for  complementary DNA  strand generation,
and  for  substring  operations  are  also included  in  this  module.
Finally,  the  \emph{alignment}-based  sequence similarity  algorithms
such  as BLAST  (using  standard  scoring matrices  such  as BLAST  or
BLOSUM) are also part of the module.

\subsection{Query System}
The data modeling  and query language for \systemname~is  based on the
ODL    and    OQL    languages,    respectively,   from    the    ODMG
standard~\cite{cattel93}.   These languages  have  been enhanced  with
support for  both the typesystems  over spatial and genetic  data, and
the operators over these typesystems.

The query processor contains,  in addition to the techniques available
in generic database systems, specialized optimization schemes for:

\begin{itemize}

\item Spatial operators, when spatial indexes are available
on predicate attributes

\item Relationship path traversals 
 
\item Queries over a type hierarchy of the data model. 

\end{itemize}

The presence of  user defined methods in the  synthesized object types
(for  example,  \emph{Print}   method  on  objects,  \emph{Area}  over
polygons, etc.), form a  serious obstacle for optimal plan generation,
since their costs  are not directly available to  the query optimizer.
A variety of strategies for handling this situation have been proposed
in  the literature~\cite{kemper91,  graefe_book}.  In  \systemname, we
have extended  the ODL language  to allow optional definition  of cost
functions,  and  functionally  equivalent methods.   These  extensions
enable the  cost-based optimizer to  compute the cost  associated with
each  of the equivalent  methods, before  choosing the  best execution
strategy.

\subsubsection*{Client Interface Framework}

The client interfacing is an important layer in the query interface of
\systemname. We have developed a simple framework to transform the
objects of the query  results into formats amenable for transportation
to  end-clients.  With  clients  following different  needs for  their
visualization  and  query  capabilities,   we  feel  this  becomes  an
important part of the query interface. Using this framework, users can
easily implement their transformation  rules which are then applied to
the appropriate objects in  the query results. The transformed results
are then shipped to the clients.

\section{Implementation Choices}
\label{sec:impl}
In this section we highlight the important software choices that we had to consider in BODHI, and provide the rationale for the decisions that we
made. We discuss these choices under the following heads: (i) selection
of storage manager and query processor, (ii) selection of access methods,
and (iii) positioning of implementation components. 

\subsection{Selection of Storage Manager and Query Processor}
For  the  back-end  storage  manager,  we selected  the  SHORE  system
developed at the University of  Wisconsin (Madison) which, at the time
we began  the project in late  1998, had a major  release the previous
year that was operational on both Solaris and Linux platforms. We were
drawn  towards  SHORE  due   to  its  attractive  array  of  features,
including:

\begin{itemize}

\item Well-implemented support for basic database functionalities such
as transactions,  logging and recovery management,  device and storage
management, etc. Recovery is implemented through the ARIES algorithm~\cite{aries}
which has become the de-facto industry standard, while multi-granularity
locking is provided for enhanced concurrency.

\item Integrates file-system interface with DBMS functionality. This can
be extremely useful in  handling genomics data which is available largely
as flat-files.

\item First-class support for user defined types. 

\item Availability of a framework for writing \emph{Value Added Servers}
(VAS) -- to provide additional features to the storage manager.

\item Presence of $R^*$-Tree~\cite{rstartree90}, a spatial indexing structure built within the SHORE kernel (in addition to the standard $B^+$-Tree index). 

\item Availability of source code, which enabled us to enhance many
of the features  of SHORE (the version we have  used is Version 1.1.1,
which was the latest at the time we began our project).

\item Successfully tested under at least two large 
scale                                                          research
prototypes~\cite{paradise94,DBLP:conf/sigmod/SeshadriP97}.

\item Intrinsic support for parallelism on a multiprocessor or network
of workstations.

\end{itemize}

After we  had been into development  for about a year,  we had reached
the stage  wherein we  were thinking about  the implementation  of the
query processor.   In particular, we were  considering the possibility
of  building our  own query  processor, using  either  a Volcano-style
framework  or  a  Tigukat-style  framework.   We  dropped  this  idea,
however, when news broke (on the \emph{dbworld}~\cite{dbworld} mailing
list) of  the first release of $\lambda$-DB,  an extensible rule-based
optimizer   from   the  University   of   Texas  (Arlington),   which,
serendipitously enough,  had been  implemented on Shore!   This vastly
reduced  our  design time  on  the  query  processor front.   Further,
$\lambda$-DB came  with an attractive set of  features including query
transformation  and optimization  rules for  OQL (specified  using the
OPTL  optimization specification  language), and  a  functional design
that made it  easy to enhance and specify  additional rules.  Finally,
it had a firm mathematical foundation in monoid comprehension calculus
that  permitted optimizations  similar  to those  found in  relational
query rewriting engines.

\subsection{Selection of Access Methods}
As  discussed earlier,  \systemname~includes  indexes for  inheritance
hierarchies,  aggregation  hierarchies,  and  spatial  data  that  are
implemented in the Object  and Spatial Services modules.
For each  of  these indexing  categories,  there have  been
numerous proposals in the research  literature, requiring us to make a
carefully selected choice.

We had intended to add  indexes for sequence  data as well.
Unfortunately, however, until this issue was addressed very recently in
\cite{ambuj2001,hunt2001}, no practical solutions for indexing the
sequences were available, rendering it impossible to realize our
objective. We are now investigating the incorporation
of these new methods in the \systemname~system.

\subsubsection*{Inheritance Hierarchies}
For indexing inheritance hierarchies, we have chosen the
\emph{Multi-key Type Indexing}\cite{mtindex97}: The basic idea behind
MT-index is a mapping algorithm that maps type hierarchies to
\emph{linearly} ordered attribute domains in such a way that each
sub-hierarchy is represented by an interval of this domain. Using this
algorithm, MT-index  incorporates the type hierarchy  structure into a
standard multi-attribute  search structure, with  the hierarchy mapped
onto one of the attribute  domains (type domain). This scheme supports
queries over single extent as well  as over extents of classes under a
subtree.   This can also  be extended  to support  the multi-attribute
queries.

Apart from its simple transformation of the tree into a linear path, a
major attraction of  the MT-index is that it  can be implemented using
any of  the multi-dimensional indexing schemes.   In particular, since
SHORE natively  supports $R^*$-trees,  the MT-index could  be directly
implemented  using this structure,  resulting in  considerably reduced
programming and integration effort.

\subsubsection*{Containment Hierarchies}
For indexing  aggregation hierarchies,  we have chosen  the \emph{Path
Dictionary  (PD) index}~\cite{pdindex98}.   The  PD-Index consists  of
three parts:  the \emph{path dictionary} which  supports the efficient
traversal of the path, and the \emph{identity index} and the
\emph{attribute index} which support associative search. The identity
index and attribute index are built on top of the path dictionary.

Conceptually,  the  path  dictionary  extracts the  compound  objects,
without the primitive attributes, to represent the connections between
these objects  in the aggregation  graph.  Since attribute  values are
not stored in  the path dictionary, it is much  faster to traverse the
nodes  in   the  extracted  path-dictionary.   In   order  to  support
associative  search  based  on  attribute  values,  PD-Index  provides
attribute indexes  which are built  for each attribute on  which there
are frequent queries.  When the  identifier of an object is given, the
path information is  obtained using the identity index  built over the
path dictionary.

On the positive side, the  PD-index supports both forward and backward
traversals of the hierarchy  with equal ease; further, its performance
evaluation    indicated    significantly    improved   access    times
in~\cite{pdindex98}.  A  limitation, however, is that  it only handles
1:1  and 1:N  relationships.  Since  typical schemas  of bio-diversity
database include aggregations of  N:M cardinality, and structures such
as sets, bags and sequences in  the aggregation path, we had to extend
the  implementation of  the  PD-index to  handle  these constructs  as
well. The details of the extensions are given in
\cite{comad2000}. 

\subsubsection*{Access Methods for Spatial Data}
For spatial data, SHORE natively supports the $R^*$-Tree~\cite{rstartree90}, which is the
most popular spatial access method since it achieves better packing of
nodes and requires fewer disk  accesses than most of the alternatives.
However, a problem  with the $R^*$-Tree is that even  though it has tight
packing to begin  with, its structure may subsequently degrade in the
presence of dynamic data.     To    tackle    this,    we    implemented
the    Hilbert
R-Tree\cite{hilberttree}, which  is designed for  handling the dynamic
spatial data while maintaining good packing of the index structure. It
makes  use  of a Hilbert  space-filling  curve  over  the data-space  to
linearize  (i.e. obtain  a  total ordering of)  the  objects in  the
multi-dimensional domain  space.  A performance evaluation
in \cite{hilberttree} shows this  structure to provide
better packing in the presence of dynamic spatial data and thus better
performance. However, the evaluation was considered in isolation and
therefore one of the goals of our study
was to investigate how well these performance improvements carried
over to a real system.

\subsection{Positioning of Implementation Components} 
\begin{figure*}
\begin{center}
\centerline{\psfig{file=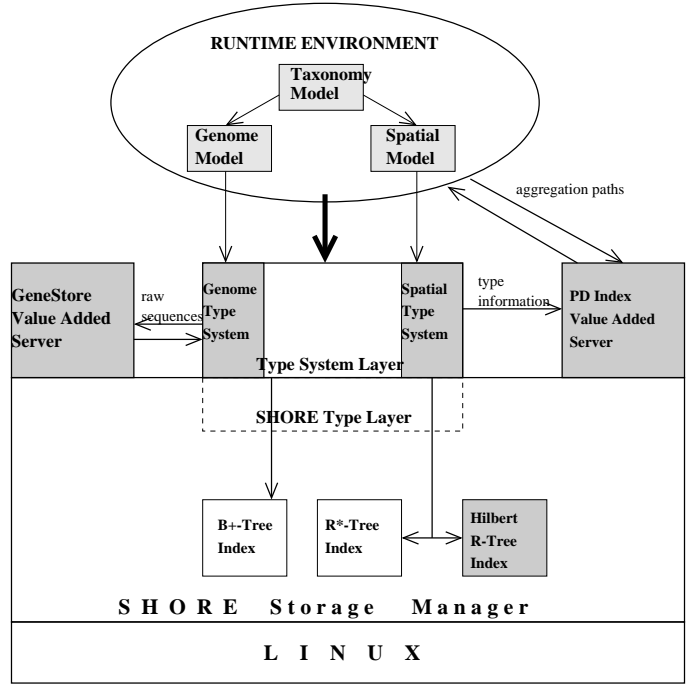,width=9cm}}
\caption{Positions of Implementation Components}
\label{fig:impl}
\end{center}
\end{figure*}
In addition to selection of software and the indexing methods, another
important  decision   that  determines  the   system  performance  and
extensibility is the placement of functionality in the implementation.
On option  is to achieve performance improvements  by supporting every
feature  of  the  system  at  the  lowest level  --  for  example,  by
implementing at the SHORE storage manager level. However, this becomes
a huge  effort to  extend and  improve the system  by addition  of new
basic  types, new access  structures, etc.   At the  same time,  if we
provide all the additional features  at layers external to the storage
manager  then the  overall  performance could  suffer.  Therefore,  we
considered these  two competing  requirements of the  system carefully
while  placing  the  implementation   of  the  services,  to  optimize
extensibility while minimizing the performance overhead on the system.

\begin{description}

\item[Object Services]{As mentioned earlier, this module bundles
the Path-dictionary and Multi-key Type indexes over object aggregation
and type  hierarchies, respectively. The  Path-dictionary structure is
implemented as  a VAS, which  maintains the path-dictionary on  a data
repository  --  with  its  own  recovery  and  logging  facilities  --
independent from the main database.  This gives the query processor an
opportunity to scan  the path-dictionary repository \emph{in parallel}
to the other data scans active at the same time.  Further, the locking
overheads are distributed over different storage management threads.

The Multi-key  Type index,  on the other  hand, is instantiated  as an
$R^*$-Tree, which  is available for  spatial indexing,   with linearized type  system as  a dimension  and each
object treated as a ``point'' in the spatial sense.}

\item[Spatial Services]{In addition to the $R^*$-Tree provided by the
Shore  storage  manager,  the  spatial services  module  provides  the
Hilbert R-Tree which  is intended for use with  highly dynamic spatial
workloads.  Although  this index  could also be  implemented as  a VAS
external to  the database,  the Shore SM  interface, the  interface to
build VASs, allows one to  introduce new logical index structures, but
no  page-level  storage  control   is  provided.   This  excludes  the
possibility of  implementing index  structures such as  Hilbert R-Tree
that rely  on physical  packing of data  for performance  benefits. We
were thus  forced to implement  the Hilbert R-Tree by  refactoring the
existing $R^*$-Tree implementation.

We had the option of implementing the spatial type system, illustrated
in  Figure~\ref{uml:spat}, either  as part  of the  basic  type system
(similar to  the support of types like  integers, strings, references,
etc.)  or at the  same level  as a  user defined  type system.  In the
former  approach, we  do gain  the storage  efficiency and  low object
creation  overhead,  but  we   lack  the  extensibility  and  ease  of
implementation available in the  latter approach. The final choice was
to go for an extensible type system, therefore, to provide the spatial
type system (along with sequence type system -- discussed below), as a
user level library which can  be modified and extended by the database
administrator without having to work on the storage manager layers.}

\item[Sequence Services]{The type system of the Sequence Services,
consisting of \emph{DNA} and \emph{Protein} types, are provided in the
same  way as  the spatial  types, which  we have  described  above. In
addition,  the  DNA  sequence  type  has extra  requirements  for  its
storage.  The  DNA  sequences   are  usually  very  long  (1000--10000
basepairs), and consists of only  4 alphabets. Instead of storing them
as character strings,  we store them in a  compressed form and perform
queries  over the  compressed  records rather  than  on the  character
strings. The efficient storage of  the raw sequences is implemented as
a separate VAS which provides advantages similar to those mentioned in
the Path-dictionary implementation.}
\end{description}

\section{Experiences}
\label{sec:experience}

In the previous sections we have described the architectural design of
\systemname~and the  specific choices  that we  made for  the various
components  of  the  design.   We   move  on  now  to  discussing  the
experiences  and  lessons  that   we  learned  during  the  course  of
implementing  these choices  in  our prototype  system.   Some of  the
issues that we  raise here with regard to  SHORE and $\lambda$-DB have
been addressed  in subsequent releases  of these code-bases --  we are
constrained, however,  to continue to  use version 1.1.1 of  Shore and
version 0.3  of $\lambda$-DB,  the versions that  were current  at the
time  we  began  the project  three  years  ago,  since we  have  made
significant  alterations and enhancements  to these  software.  

\begin{figure*}[tb]
\centerline{\psfig{file=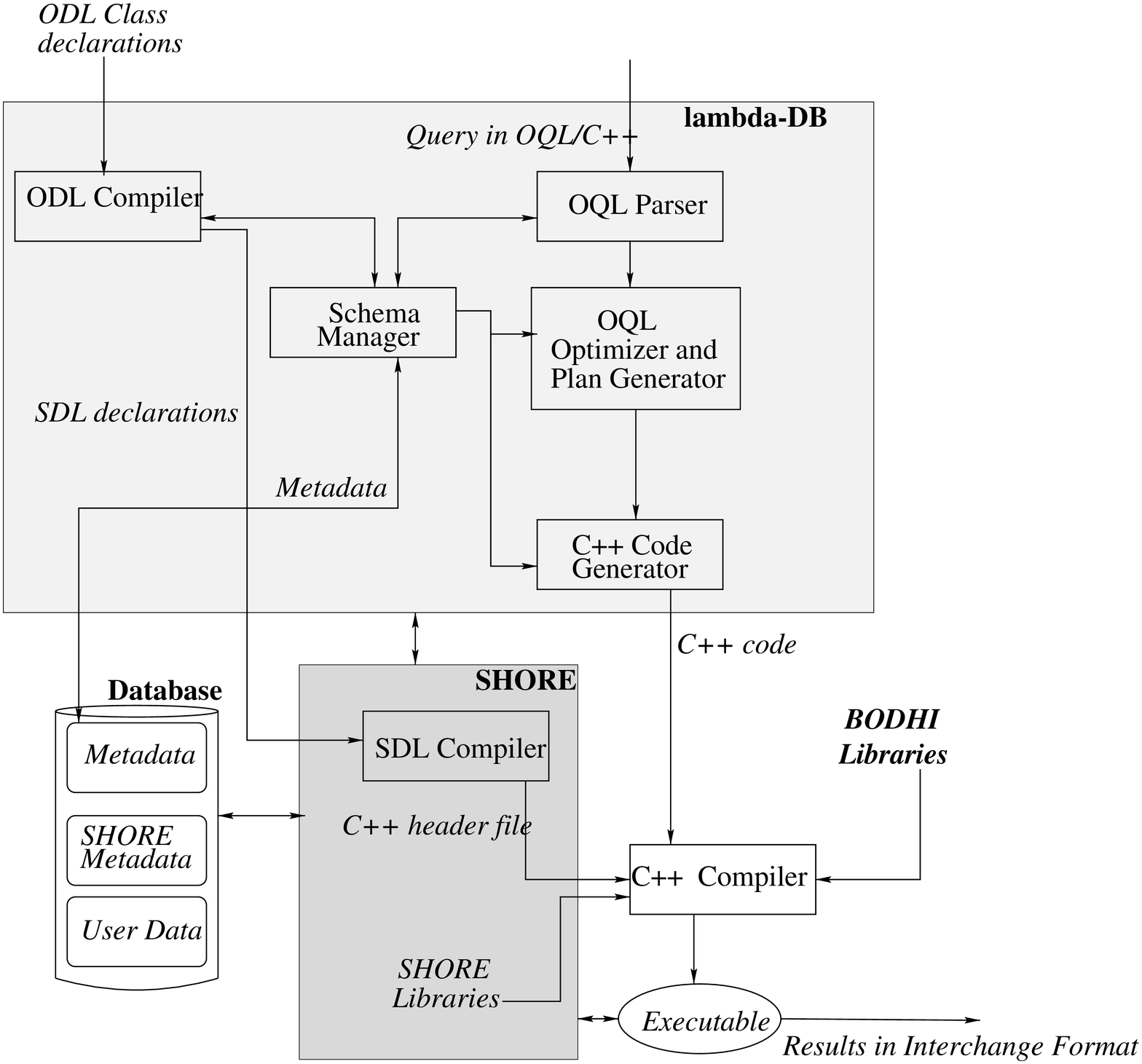,height=10cm}} \caption{Schema
Definition and Query Flow in \systemname} \label{fig:qpflow}
\end{figure*}

The overall  detailed implementation of  the system is  illustrated in
Figure~\ref{fig:qpflow}.  As illustrated,  the schema  declarations in
ODL are first converted into  SDL (the definition language provided on
top   of   the   SHORE   storage  manager),   by   $\lambda$-DB.   The
implementations of  the schema declarations  are stored in  a separate
source  file  that  is  compiled  into  a  linkable  library  for  the
applications. Similarly,  the query in the OQL  format is typechecked,
optimized  and converted  into an  implementation of  the optimal physical
plan by $\lambda$-DB.

\subsection{Index Key Formats}
$\lambda$-DB generates the query implementation making use of its
runtime interface to the SDL layer of SHORE. The query is evaluated in
a streaming fashion, avoiding the materialization of the sub-queries as
much as possible. Indexing over object extents is achieved by maintaining
a separate extent of indexes. In SHORE, the index objects have to
reside within a ``user level'' object. Now, while $\lambda$-DB uses an
\emph{ExtentIndex} type to hold the indexes, it also converts all the
index keys  into a \emph{string} format  in order to handle  them in a
generic  way.  This turns  out be  a problem  when handling  keys that
cannot be  converted into  character strings (such  as in the  case of
spatial  indexes), and  in handling  keys which  result in  a  loss of
information  during the  conversion (such  as floating  point values).
Therefore, in  order to support  the spatial indexes from  the ODL/OQL
layers, we were forced to introduce a specialized key type for spatial
indexes and also implement a special index holder class. This required
a considerable  amount of modification  and extensions to the  code in
the query processor.

At the same time, the rule-based optimization scheme of the
$\lambda$-DB simplified the process of adding new operators into OQL, as well as their optimization and rewritings into the physical operators
based on the statistics. We added operators such as \emph{Overlaps},
\emph{Inside} etc. for spatial operations, and sequence retrieval
operators such as \emph{BLAST} into the OQL specification supported by
the query processor. 

\subsection{Index Visibility}
The implementation  of access structures  for spatial data  and object
hierarchies  raised some of  the  subtle issues  with regard  to
hosting them  on the  $\lambda$-DB and SHORE  combination. One  of the
most surprising revelations  was the lack of spatial  index support at
the SDL  layer in SHORE  -- which is  still not available  since there
have  been no further  releases of  the SDL  layer. The  $R^*$-Tree is
available only  at the storage manager  level, but is  not exported to
the SDL interface.   This also meant that $\lambda$-DB  which uses the
SHORE through the  SDL interface also has no  knowledge of the spatial
indexes. In order to provide the support at the OQL level we first had
to  rework  the SHORE  code,  and then  integrate  it  with the  query
processor.

\subsection{PD Index Implementation}
While implementing the  Path-dictionary-based indexing for aggregation
path  queries, we  found  that  the index  structure  as presented  in
\cite{pdindex98} cannot be used in a stream based query processor such
as   $\lambda$-DB,  without  breaking   the  pipeline   structure  and
materializing the query  results at that join node.  We addressed this
problem by \emph{inverting}  the storage of paths to  proceed from the
top  of  the  aggregation  tree  instead of  the  suggested  bottom-up
approach.  While this inversion may partially reduce the effectiveness
of the path-dictionary, the major benefit of avoiding the huge cost of
joins over object extents is retained.

\subsection{VAS Feature}
In  building the  PD-index, we  exploited the  concept of  Value Added
Server  (VAS), one of  the strong  features of  SHORE. The  ability to
provide  a concurrent  storage manager  with  a full  set of  database
features such as transactions,  logging, recovery etc., eased the task
of extending  the storage manager  capabilities tremendously. Although
RPC-based interaction between the  storage server instances results in
communication  delays  and  reduced  type-support across  the  storage
servers,  it enables cleaner  separation of  services provided  by the
storage manager. 

We also used  the VAS feature to provide  genome sequence storage, and
retrieval algorithms over this storage. An important advantage of this
implementation is that it is  easy to extend and optimize the sequence
retrieval algorithms  without having side-effects  on the rest  of the
system. A problem, however,  was the following: The storage allocation
of the sequences  on the VAS is effected  through a specific interface
which stores  the sequences in a  compressed form on  the disk. Though
this storage should  ideally be handled transparently, due  to lack of
\emph{post-construction    hooks}   for   object    instantiation   in
$\lambda$-DB and  SHORE, this compressed  storage of sequences  has to
be explicitly called during database loading.

\section{Experimental Results}
\label{sec:expt}

We have  evaluated the  performance of BODHI  on a testbed  of typical
queries  in the  bio-diversity domain.  These  queries make  use of  a
mixture of  synthetic and  real datasets and  consist of  queries over
both \emph{single-domain} (such as taxonomy,  spatial or sequence domains) and
\emph{multiple     domains}      --     i.e.,     queries      similar     to
Query~\ref{query:grand} in the Introduction.  Moreover, since spatial  data forms  a large
fraction of data and is traditionally considered the main component of
the query processing  time, we studied the performance  of the spatial
component  in detail.  In  particular, we  evaluated the  spatial data
handling capabilities  of BODHI over  the datasets and queries  of the
\emph{Sequoia  2000} regional benchmark\cite{sequoia2000},  a standard
benchmark for spatial databases.

The  performance  numbers reported  were  generated  on a  Pentium-III
700MHz  processor,  with  512MB  memory  and an  18GB  10000-RPM  SCSI
harddisk (IBM  DDYS-T18350M model), connected  with Adaptec AIC-7896/7
Ultra2 SCSI host  adapter.  In order to reduce  the effects of Linux's
aggressive memory mapping of  files, we flushed the benchmark database
each time with an I/O over a large database. Further, we kept the size of the buffer pool used by the storage manager to just 320KB (corresponding setting in Paradise is 16MB).

In rest of the section,  we first describe the synthetic datasets used
in our queries,  and then proceed to report  the capabilities achieved
and performance results of representative queries over BODHI.

\subsection{Description of Datasets}

\begin{table}
\renewcommand{\baselinestretch}{1}
\normalsize
\begin{center}
\begin{tabular}{|l|c|}
\hline
Parameter&Value\\
\hline\hline
Branch  Factor  at  each  level  of Taxonomy  &  U(1,19)\\\hline  Mean
(height, width) of habitat regions &(10,12)\\\hline
\multirow{4}{200pt}{Range of distribution of habitat regions}
&from \\  &(-100, -100)\\ &to  \\ &(-1000, -1000)\\\hline No.   of DNA
sequences per species&10\\\hline
\end{tabular}
\caption{Parameters to Synthetic Data Generator}
\label{tab:synthparam}
\end{center}
\end{table}

\begin{table}
\renewcommand{\baselinestretch}{1}
\normalsize
\begin{center}
\begin{tabular}{|l|l|l|}
\hline
Element &No. of Tuples &Overall Size(in KB)\\
\hline \hline
Order &4 &0.6 \\ Family &46 &7.1 \\ Genera &496 &76.0 \\ Species &5155
&1153.1 \\ FlowerChar &5155 &564.0 \\ Habitats &5155 &607.0
\\ InfloChar &5 &20.4 \\ EMBLEntry &51550 &2902 \\
\hline \hline
Total & &5330.2 \\
\hline
\end{tabular}
\caption{Statistics of the Synthetic Dataset}
\label{tab:synthdata}
\end{center}
\end{table}


The synthetic data used in  our experiments conforms to a biodiversity
object  model, which  is presented  in part  as an  object  diagram in
Figure~\ref{fig:model}.   Even   though  we  collaborated
closely with  the scientists of  the ecological sciences  in designing
this  object   model  to   represent  their  requirements,   we  faced
difficulties in  procuring enough  data to be  used in  the evaluation
experiments of  the system. This is because, the domain  experts have a bulk of their
data in legacy formats, and often on ``herbarium sheets''\footnote{These are sheets that contain
  a plant specimen and its details.} they maintain.
We are currently in the process of converting this data for populating
the database,  and have therefore  chosen readily available  subset of
real data, and boosted the dataset with synthetic data.

As shown in the object model, the schema is hierarchical in nature and
consists  of  aggregation paths,  inheritance  structures over  object
types, spatial and genome sequence components. The well known taxonomy
aggregation path of  Order-Family-Genera-Species forms the backbone of
the  model.   Each  Species   has  a  set  of  identifying  characters
(IdentChar), and there are many sub-characteristics that are inherited
from this. The spatial component of the model consists of a collection
of reported habitat areas for  each Species. Also associated with each
Species is a  collection of DNA sequences that are  used to study the
evolutionary    pathways    amongst    the   species    by    locating
homologies (sequences which  have high  likelihood of sharing  a common
ancestor). We now describe  the mechanism of generating synthetic data
which complies to the object model.
 
\begin{description} \item[Taxonomy Data]
{   We   generated   the   object  relationships   in   taxonomy   and
characteristics  hierarchy  by  setting  a  heuristic  probability  of
association at each optional  relationship.  In case of collections in
the  aggregation  path,  the  branch  factor  of  the  collection  was
uniformly  distributed  over   specified  end-points.  The  real  data
available for about fifteen  closely studied Plant species was boosted
with this synthetic data.}

\item[Spatial Data]
{We used the synthetic data generation method followed in
\cite{hilberttree}.  The data consists of rectangular regions, whose
centers  are uniformly distributed  over a  unit square.   The overlap
between  rectangular  regions  can  be controlled  by  specifying  the
distribution of their height and width values. It should be noted that
this dataset consists of only rectangular regions, while in reality we
have to handle  non-convex polygonal regions as well. The performance of spatial data handling over real dataset (involving non-convex polygonal regions) will
be evaluated separately through the  Sequoia 2000 benchmark.  Each species object
generated above  is associated with a  synthetically generated polygon
that represents the habitat of the species.}

\item[Genome Data]
{In the  case of Genome sequence data,  we could use the  data that is
publicly  available through repositories  such as  GenBank, SwissProt,
etc.  In our experiments, we made use of a randomly selected sample of
``expressed sequence  tags'' (ESTs) of various genomes  available from the
BLAST  database  of  EMBL  GenBank~\cite{www:genbank}.  We  used  these
sequences  to  populate  the  DNA  information  of  our  synthetically
generated species.}

\end{description}
\noindent
We summarize the parameters used for the benchmark dataset in Table~\ref{tab:synthdata}. We consider a set of 5 queries, over this dataset that conforms to the schema illustrated in Figure~\ref{fig:model}. These queries span the domains of taxonomy, spatial and genome data, and illustrate the capabilities of BODHI in handling these domains. In addition, the performance numbers of these queries provide an indicator towards overall expected performance of the system.

\subsection{Biodiversity Queries}

We  now describe  the  set  of queries  considered  to illustrate  the
capabilities  of BODHI  and present  the performance  numbers over
each  of these  queries. The  query mix  can be  split further  into 3
categories: Taxonomy queries, Genome queries and Multi-Domain Queries.
We collectively refer to Taxonomy and Genome Queries as
\emph{Single-domain queries}, since predicates involve either
taxonomy hierarchy or genetic  sequences associated with a species, but
not both.  The Multi-domain queries,  on the other hand,  query across
taxonomy hierarchy, habitat (spatial) collection and genetic sequences
data corresponding to species. The performance numbers for the queries
are  summarized in  Table~\ref{tab:SDQ} and  Table~\ref{tab:MDQ}.

\begin{center}
\begin{table}
\begin{center}
\renewcommand{\baselinestretch}{1} 
\normalsize
\begin{tabular}{|l|l|} 
	\hline    \bf    Id     &\bf    Time    \\    \hline    \hline
	\multirow{2}{60pt}{\bf Taxonomy  Query-1} & 73  min.  (Without
	Pathdictionary)   \\    \cline{2-2}   &   0.5    min.    (With
	Pathdictionary) \\ \hline \bf Genome Query-1& 0.2 sec.\\\hline
	\bf Genome Query-2& 1.5 min.\\ \hline
\end{tabular}
\caption{Performance Numbers for Single-domain queries}
\label{tab:SDQ} 
\end{center}
\end{table}
\end{center}

\begin{center}
\begin{table}
\begin{center}
\renewcommand{\baselinestretch}{1} 
\normalsize
\begin{tabular}{|l|l|l|l|}
\hline
\bf Id & \bf Without Index&\bf Pathdictionary &\bf Spatial \&\\ 
& & &\bf Pathdictionary \\ \hline\hline
\bf MDQ1 &26.99 &11.13 &2.1 \\ \hline
\bf MDQ2 &553.66 &542.12 &530.2 \\ 
\hline
\end{tabular}
\caption{Performance Numbers for Multi-domain Queries}
\label{tab:MDQ} 
\end{center}
\end{table}
\end{center}

\begin{description}
\item[Taxonomy Query-1]
{\emph{Find all species that have the same Inflorescence characteristic
in their Flowers as that  of ``Magnolia-champa''.\\} This query performs a
three  level path  traversal  over aggregation  hierarchy of  Species,
Flower  and  Inflorescence  Characteristics.    By  referring  to Table~\ref{tab:SDQ},  we see  that
without any indexing strategy  for accessing the aggregation paths, the
query execution times are  unacceptably high -- especially considering
the  modest  size  of  the  dataset.  The  performance  of  the  query
execution improves by  two orders of magnitude with  the presence of a
Pathdictionary index  over the queried path.  As  discussed earlier in
Section~\ref{sec:impl},   the  Pathdictionary   maintains   a  compact
materialization of  joins along queried path,  preventing the repeated
computation of these expensive  joins. Interestingly, if we follow the
aggregation paths through the usage of ``swizzled pointers'' available
through  C++ interface of  SHORE, this  query can  be answered  in 8.5
seconds, which  is much faster  than even using  Path-Dictionary based
indexing.   It  has  to   be  noted   that  rewritings   available  in
query-processors  such  as  $\lambda$-DB  do  not make  use  of  these
features  available with  the storage  managers, thus  incurring heavy
cost of joins.}
 
\item[Genome Query-1: ] 
{\emph{Retrieve  all  DNA  sequences  of Magnolia-champa.\\}  The  DNA
sequences  are  stored  encoded,  using context-free  encoding,  in  a
separate  storage. This encoding  increases the  disk-memory bandwidth
and  enables the  sequence similarity  algorithms to  operate  in this
encoded  domain itself.  At the  same time,  there is  an  overhead of
decoding them  before presenting to the user.  The performance numbers
of this  query give estimate of  the delay involved  in decoding these
sequences}

\item[Genome Query-2: ] 
{\emph{List names of all Species  that have a DNA sequence  within a BLAST
score of 70 with any sequence of Magnolia-champa.\\}
The  computation of  BLAST  scores over  a  database could  be a  time
consuming task. We don't have  any indexing schemes for speeding these
queries,  for reasons  mentioned earlier  in Section~\ref{sec:design},
and hence for each  query sequence we have to make a  full scan of the
sequence  database  and  compute   the  scores,  significance  of  the
alignments, etc.   The timings for this  query -- which  results in 10
BLAST  computations,  is  given  in Table~\ref{tab:SDQ}.   When  these
numbers are compared against the query capabilities of BLAST-farms run
by organizations such as EMBL, they might look rather high.  However,
these  BLAST-computation farms  make  use of  large-scale and  heavily
optimized  data handling  equipment and  keep the  entire  database in
memory for speeding  up the processing times, while  BODHI is aimed to
handle varied  data, and is  running on a general  purpose small-scale
machine.}

\item[Multi-domain Query-1: ]
{\emph{Find  all Species  sharing  a common  habitat  and having  the same
Inflorescence  characteristic  as  Magnolia-Champa.\\}  This  query  is
targeted at  the combination of hierarchical data  of Taxonomy domain,
and associated Spatial data.  Being  one of the popular queries by the
collaborating ecologists, this evaluates the combined effectiveness of
Pathdictionary index  and $R^*$-Tree  indexes available in  BODHI. The
performance  numbers  provided  in  Table~\ref{tab:MDQ}  are  for  the
optimal query plan which performs the spatial overlap before computing
the joins over the aggregation  paths. Since spatial overlap is highly
selective  in the  existing dataset,  the number  of  path aggregation
traversals are  reduced to a very  small number.  As a  result, we see
that even  though this query  is more complicated  than \emph{Taxonomy
Query-1}, it takes less than 0.6\% of time taken for
\emph{Taxonomy Query-1} even in the absence of Path-Dictionary index. 
The presence  of Path-Dictionary  reduces the execution  time further,
from 26.99  seconds to 11.13 seconds  -- a reduction  of 58\%. In
this case,  the execution times  are dominated by the  spatial overlap
computation.  We can see this clearly by looking at the performance of
the query,  when both $R^*$-Tree and Path-Dictionary  are present. The
query   time   is   just   around   2   seconds,   almost   80\%
 improvement. This clearly indicates  that both indexing strategies are
extremely useful for such queries. }

\item[Multi-domain Query-2: ]
{\emph{Retrieve  all Species  sharing  a common  habitat, having  same
Inflorescence characteristics and having a  DNA sequence within BLAST score of
80, with respect to Magnolia-champa.\\}This query, which extends the
\emph{Multi-domain Query-1} by adding extra predicate for BLAST score
computation  for each  of  the  sequences in  the  target species,  is
equivalent  to  the  ``goal''  query  that  we  presented  earlier  as
Query~\ref{query:grand} in the Introduction.  This query is written in the BODHI system, using OQL,as follows:\\

\begin{minipage}{6.0in}
\renewcommand{\baselinestretch}{0.9}
\small
\begin{verbatim}
select * from species1 in PlantSpecies, 
              species2 in PlantSpecies,
              embl1 in species1.stDNAEntries, 
              embl2 in species2.stDNAEntries 
        where
              species1.flowerchar.inflochar = species2.flowerchar.inflochar
              and
              species1.georegion overlaps species2.georegion
              and
              embl1 in embl2.dna.blast(80);
\end{verbatim}
\vspace{0.1in}
\end{minipage}

Referring to Table~\ref{tab:MDQ}, we  see that the execution times are
much higher than those of
\emph{Multi-domain Query-1} -- due to the additional 50 BLAST computations. 
The reduction in execution times are approximately same as in
\emph{Multi-domain Query-1}, about 11 seconds in presence
of  Path-Dictionary  and by 10  seconds  in
presence of  both $R^*$-Tree and Path-Dictionary  indices. Hence, this
query is clearly dominated by the BLAST computations.  Therefore, it appears to be imperative to develop indexing strategies to improve performance of such queries over genome sequence data.}
\end{description}

\subsection{Evaluating Spatial Data Handling}
The  evaluation of queries  over spatial  data has  traditionally been
considered as a highly  compute-intensive operation, and many indexing
strategies  have been  proposed to  improve the  performance  of these
queries. The  SEQUOIA benchmark has been quite  popular for evaluating
the performance and capabilities  of spatial databases. It consists of
a set of 10 queries over  a schema involving the spatial objects (such
as polygons, points and graphs) and  also bitmap (raster) objects. As we do not
have  support for  bitmap data  formats in  BODHI, we  have  chosen to
ignore  the  raster dataset  and the queries  (2),(3),(4) \and (9),  which involve
these objects.  The  vector benchmark  data consists  of 62556
Point  objects,   58585  Polygons  and  201659   Graph  objects.   The
Table~\ref{tab:sequoia}  summarizes  the  performance numbers  of  the
queries.   We have  compared our  performance numbers  with  a spatial
database  system,  Paradise~\cite{paradise94},  also  built  on  SHORE
storage   manager,  and  Postgres~\cite{postgres},   a  successful   and  free
object-relational database.   The numbers given for  these two systems
are taken from those reported in~\cite{paradise94}. Note that overall, performance numbers of BODHI are very similar to those of Paradise.

\begin{table}[htb] 
\renewcommand{\baselinestretch}{1} 
\normalsize
\begin{center}
\begin{tabular}{|l||l|l|l|} 
	\hline \bf  Id &\bf BODHI &\bf  Paradise&\bf Postgres\\ \hline
	\hline \bf  1&5742.0&3613.0&8687.0\\ \bf 5 &0.12  &0.2 &0.9 \\
	\bf 6 &8.0 &7.0 &36.0 \\ \bf  7 &0.66 &0.6 &30.5 \\ \bf 8 &9.7
	&9.4  &62.2 \\ \bf  10 &11.0  &\emph{Not supported}  &327.2 \\
	\hline
\end{tabular}
\caption{SEQUOIA Benchmark numbers}
\label{tab:sequoia} \end{center} \end{table}

We now  briefly explain  the chosen set  of SEQUOIA queries  and their
performance statistics  with reference to  Table~\ref{tab:sequoia}. We
also note the importance of some  of these in a typical set of queries
expected in bio-diversity studies.

\begin{description} 
\item[Sequoia 1 -- Dataloading and indexing.]{This query populates the
database from  a given set of  datafiles, and is  expected to exercise
the bulk-loading facility in the database.  At the time of writing, we
still do  not have the bulk-loading  feature in BODHI,  resulting in a
transaction commit  for each  object hierarchy.  Therefore,  the table
represents only an upper bound  on the dataload and indexing times for
the  spatial component. Referring  to Table~\ref{tab:sequoia},  we see
that this is the only benchmark query in which BODHI is far worse than
Paradise which  supports bulkloading facility.  However,  we don't see
it as a  major bottleneck in BODHI, since  the bio-diversity databases
are not  expected to  have high rates  of bulk data  updates. Instead,
these databases  are highly query-intensive and hence  it is important
to  have  fast  query  processing  speeds.   In  addition,  we  expect
improvements in performance when the bulk-loading scheme is put
in place for BODHI.}

\item[Sequoia 5 -- Select a point based on its name.]{The performance of
the  B-Tree index  over the  name of  the point  is evaluated  in this
query.  We  see from  the table that  both BODHI and  Paradise perform
fairly  well on  this query  -- which  is expected  due to  the common
storage manager, i.e. SHORE, used in both systems.}

\item[Sequoia 6 -- Select polygons overlapping a specified rectangle.] 
{This  is one  of  the  typical spatial  queries  asked in  ecological
studies where a geographic region is split into a set of grids and the
researchers  would  want  to  identify the  species  whose  previously
recorded habitat boundaries overlap  with the grid being studied. This
could  be important  in identifying  species whose  co-existence  in a
region  is to  be  targeted  for study.   The  performance of  spatial
operators such as \emph{overlap} depend directly on the performance of
implementing these operators on an spatial index such as $R^*$-Tree or
Hilbert R-Tree.   Since the $R^*$-Tree implementation of  BODHI is the
same as that of Paradise (both use the index provided by the SHORE storage
manager), we don't see much difference in the query execution performance.}

\item[Sequoia 7 -- Select polygons greater than specified area,
contained  within a  circle.]   {  We see  similar  query occuring  in
bio-diversity studies with variations  in the area selection clause of
the  query.  The  area  of a  polygon  is provided  through a  derived
attribute -- computed  based on the co-ordinates of  the polygon. This
is extendible to allow for selection over arbitrary derived attributes
over which an index can be built. Thus, in ecological study databases,
we get variations  of the query that locate all  the habitats that are
near a study center, with  a derived attribute value (such as bio-mass
index of the habitat, etc.).

This query reflects the combination  of B-Tree and spatial index based
query  processing. The  order in  which this  query gets  evaluated --
whether the B-Tree lookup or the $R^*$-Tree based overlap selection is
made  as  the first  step  --  makes a  big  difference  in the  query
answering times.   The usage of  query optimizer which  maintains cost
statistics and uses it to arrive at the final evaluation order is also
tested    in     this    query.     The     numbers    presented    in
Table~\ref{tab:sequoia} are  for the  optimal plan generated  by query
processor of BODHI,  which is to perform the  $R^*$-Tree based overlap
selection first and then the B-Tree-based polygon area selection.}

\item[Sequoia 8 -- Select polygons overlapping a rectangular region
around a point.]  {This query  involves a spatial join between a point
and polygons.   When scientists are interested in  regions around test
points satisfying a  criterion in their study area,  this is the query
they would be using to derive the information.}

\item[Sequoia 10 -- Select points contained in polygons with specific
landuse  type.]  {This  query is  a  join between  polygon extent  and
points through an
\emph{inside} predicate.  The SEQUOIA benchmark specifies
\emph{islands} within polygon regions. In order to get the right set
of  answers,  we  should  exclude  the  points  which  fall  in  these
islands(or ``holes'')  in the polygons.  The Paradise  system does not
report   numbers  for   this  query,   since  it   does   not  support
the~\emph{minus}  operator \cite{paradise94}.   In  BODHI, we  perform
this operation using a subquery which implements the
\emph{minus} operation. 

The ecological  data gathering is done at  various specimen collection
centers  in  an ecoregion.   The  ecoregions  are  usually split  into
different  forest  types and  researchers  are  usually interested  in
locating  collection  points  which  are located  in  specific  forest
type(s). At  the same time, she  might want to  exclude the collection
points for the aquatic organisms.}

\end{description}

\begin{table}[htb] 
\renewcommand{\baselinestretch}{1} 
\normalsize
\begin{center}
\begin{tabular}{|l||l|} 
	\hline \bf  Id &\bf with Hilbert R-Tree \\ \hline
	\hline \bf  1&9798.0\\
	       \bf  5&0.12\\
	       \bf  6&60\\ 
	       \bf  7&0.4\\ 
               \bf  8&56.0\\ 
	       \bf  10&14.0\\
	\hline
\end{tabular}
\caption{SEQUOIA Benchmark numbers with Hilbert R-Tree}
\label{tab:hilbert} \end{center} \end{table}

We also  executed these Sequoia benchmark queries  with Hilbert R-Tree
in  place  of  $R^*$-Tree.  The  results obtained  are  shown  in  the
Table~\ref{tab:hilbert}. It was surprising to see that the performance
of these queries  in presence of Hilbert R-Tree  is considerably worse
than  with  $R^*$-Tree.  Our  initial  investigation  into  the  cause
indicated that the  packing factor of Hilbert R-Tree  is extremely low
(less  than   half  of  $R^*$-Tree),  resulting  in   many  more  disk
accesses. We are currently  studying our implementation of the Hilbert
R-Tree closely to ascertain our initial observation for this behaviour.

In addition, BODHI also supports the spatial aggregate operator
\emph{Closest}, on the lines of Paradise spatial data management
system.   This operator  was used  in executing  three of  the spatial
aggregate  benchmark queries  given by  Paradise system,  as Query-11,
Query-12  of   their  benchmark   queries   reported
in~\cite{paradise97}.  For completeness, we have also included Query-13, which is not an aggregation query, but is a spatial join in benchmark queries of Paradise.  But we cannot  compare the  performance numbers
obtained  in BODHI with  those reported  in \cite{paradise97},  as the
benchmark  datasets are  completely different  in both  schema  and the
scale (they used 10 years of 8 Km. resolution AVHRR satellite images obtained from NASA, and DCW global data set containing information about roads, cities, land use, drainage properties etc.).  Hence, we   present  the   numbers   in   an   absolute  sense   in
Table~\ref{tab:paradise}.
\begin{table}
\renewcommand{\baselinestretch}{1} 
\normalsize
\begin{center}
\begin{tabular}{|l||l|} 
	\hline \bf  Id&\bf Time  \\ \hline
	\hline 
	\bf  11&3.36 sec.\\ 
	\bf  12&51 sec.\\
	\bf  13&66 min.\\ 
	\hline
\end{tabular}
\caption{Performance over Paradise Queries}
\label{tab:paradise} \end{center} \end{table}

\begin{description} \item[Paradise 11 -  Select closest graphs (polylines) to a
given  point.]  {This  query requires  the evaluation  of  the spatial
aggregate   ``Closest''  using   available  index   structures.   This
aggregation operator is implemented  as an iterative searching for the
closest polyline (\emph{Graph} in Sequoia dataset).  At each iteration step, a box is constructed around
the  given point,  and all  polylines that  overlap with  the  box are
located  using the  spatial index.  If no  polyline that  satisfy this
constraint are  found, then  dimensions of the  box are  increased and
another  iterative search is  performed.  When  we obtain  a non-null
candidate set  through this step,  we compute exact  distances between
the point and the polylines in the candidate set, and closest polyline
is determined.  The  performance of this query depends  heavily on the
location  of  the point  and  the  distribution  of polylines  in  the
region.  If the  polylines  are  densely packed,  we  get to  non-null
candidate  set within  a  few  iterations (most  likely  in the  first
iteration    itself),    thus     getting    the    target    polyline
instantaneously.     The    performance    numbers     presented    in
Table~\ref{tab:paradise}  were obtained  over a  sample of  100 points
from Sequoia dataset.}

\item[Paradise 12 - Select closest graphs to every point.] 
{This query is  an extension of earlier query,  and performs a spatial
aggregate on a cross product  of two relations (in this case polylines
and points). For each point in the Sequoia dataset, we use the \emph{Paradise 11} presented earlier, and locate the closest polyline.}

\item[Paradise 13 - Select all polylines which intersect with each other.] 
{This query joins two large spatial relations and tests the efficiency
of  the  system's  spatial  join  algorithm. The  cardinality  of  the
polyline extents in Sequoia benchmark  is very high, with 201659 graph
objects in  the dataset.  In order  to answer this  query, we  need to
perform a self spatial-join of this extent, which is highly expensive.
This is  clear from 66 minutes  reported in Table~\ref{tab:paradise},
to answer this query.  }

\end{description}

The SEQUOIA benchmarks show that BODHI is very close in performance to
the  performance  of  Paradise, a specialized and highly optimized spatial  database
system. Even though the hardware platform used by both the systems are
difficult to compare, it should be noted that both Paradise and BODHI use
the  same storage manager  system.  In  addition the  following points
regarding  numbers  reported under  BODHI  should  be  noted: (i) We use file-based storage management instead  of  using raw-disk  as  done  by  Paradise system.   (ii)The
optimal   physical  query   plan  is   generated  through   a  generic
object-oriented  query   processor.   (iii)The  typesystem   is  easily
extendible as it is provided as an ODL schema defintion.

\section{Related Work}
\label{sec:related}

Bio-diversity data consists of both macro-level and micro-level
information ranging from ecological information to genetic makeup of
organisms and plants. Apart from our work, we are not aware of any
other that attempts to combine the complete spectrum of information,
though the need for it is highlighted in a recent proposal for
\emph{GBIF}(Global Bio-diversity Information Facility)\cite{oecd:gbif}
by OECD(Organization for Economic Co-operation and Development). This
proposal identifies the domain level challenges in building a global,
interconnected data repository of bio-diversity information systems and
notes that the urgent requirement in bio-diversity studies is a
suitable information management architecture for handling vast amounts
of diverse data.

In the area of macro-level bio-diversity data management, there have
been many governmental efforts from various countries such as
\emph{ERIN}\cite{erin:www2}, \emph{INBio}\cite{inbio} and global
initiatives such as \emph{Species 2000}\cite{species2000}, \emph{the Tree
of Life}\cite{treeoflife98}, etc. And in a recent
report sponsored by the National Science Foundation SF in
the USA \cite{nsf-usgs:2001}, a group of computer scientists 
have outlined research directions in bio-informatics.

The micro-level bio-diversity data, or genetic information of various
species, has been growing steadily due to a multitude of genome
sequencing initiatives. The specific data management issues in handling
such data\cite{goodman:labbase, goodman:challenges} have been addressed
in quite a few proposals. In all of these proposals, the database
management architecture has been tailored for the specific purposes of
the project. For example, consider the \emph{ACeDB} (A C.elegans
Database)\cite{acedb} database system, originally proposed for \emph{C.
elegans} genome sequencing project. ACeDB is an object oriented data
management tool that has many features that make it a extremely popular
software in many sequencing projects. ACeDB handles missing data and
schema evolution issues, common requirements in a ongoing sequencing
projects, in a flexible manner. However, in spite of its popularity in
genome sequencing community, it cannot be considered for the larger
requirements of bio-diversity data handling due to the following
reasons: (1) Its lack of support for geo-spatial data; (2) Weak support
for database updates; and (3) The  lack of recovery mechanisms
necessary in large data repositories.

In \systemname, we have provided the key strengths of ACeDB (its
sequencing algorithms and object-oriented basis), and augmented it with
strong database functionalities, in addition to other features
necessary for a complete bio-diversity information repository.


\section{Conclusions}
\label{sec:concl}
We have reported in this paper on our experiences in building BODHI,
an object-oriented database system intended for use in bio-diversity
applications. To the best of our knowledge, BODHI is the first system
to provide an integrated view from the molecular to the organism-level
information, including taxonomic data, spatial layouts and genomic
sequences. 

BODHI is operational, completely free and is built around publicly
available software components and commodity hardware.  Further,
BODHI incorporates a variety of indexing strategies taken from the recent
research literature for efficient access of different data types. Through
a detailed performance study using a range of biological queries, we
showed that these indexes were extremely effective in reducing the running
times of the queries.  Our experiments also showed that while spatial
operations are certainly expensive, as mentioned in the literature, in
the biological context, it is perhaps the genomic sequencing queries that
are really the ``hard nuts''.  Therefore, the importance of developing
efficient indexing strategies for sequence data cannot be over-emphasized.

We hope that BODHI can be successfully used by biologists as the central
information repository of their workbench, and by computer scientists
as a realistic testbed for evaluating the efficacy of their algorithms.
We are currently working on adding new indexing mechanisms such as
the Pyramid Tree for indexing high-dimensional data, where each data
object has thousands of attributes -- such data is especially common in
drug-related datasets.

\bibliographystyle{plain}

\end{document}